\documentstyle[prl,aps,twocolumn,epsf]{revtex}
\def\break#1{\pagebreak \vspace*{#1}}
\begin{document}

\draft

\title{Comment on quant-ph/9509008 by Kumar and Khare}

\author{H.C. Rosu\cite{byline}
}

\address{ %$^{a}$
{Instituto de F\'{\i}sica de la Universidad de Guanajuato, Apdo Postal
E-143, L\'eon, Gto, M\'exico}
}

%\date{submitted to ABCDE, `prin avion', some mmddyy}
\maketitle
\widetext

\begin{abstract}
One can find some comments related to the isospectral issue.

\end{abstract}

\pacs{PACS numbers:  11.30Pb  \hfill
LAA number: quant-ph/9608017}

%%%%%%%%%%%%%%%%% THE COMMENT/PROPERTY OF THE HUMAN MIND %%%%%%%%%%%%%%%
%%%%%%%%%%%%%%%%%%%%%%%%% written by H.C. Rosu  %%%%%%%%%%%%%%%%%%%%%%%%%%%%%
%%%%%%%%%%%%%%%%%%%%%%%%%    August 11, 1996  %%%%%%%%%%%%%%%%%%%%%%%%%%%%%

\narrowtext
Recently, Kumar and Khare \cite{kk} constructed in a general manner
coherent states for
strictly isospectral Hamiltonians $H_{\lambda}$, where $\lambda$ is the
isospectral family parameter, by using a unitary transformation relating
$H_{\lambda}$ to the original Hamiltonian $H_{|\lambda|\rightarrow \infty}$.
This is different from old proposals regarding the isospectrality issue
\cite{m,n},
and appears as the natural construction for the discrete part of the
spectrum. This is why, Kumar and Khare
commented in the negative on the old results on this problem.
However, without questioning the good merits of their work,
I have found the remarks they did relative to \cite{m} as too negative,
though perhaps some statements in \cite{f3} may be judged as incorrect.
Therefore, I decided to write this comment.

Mielnik's ``annihilation" and ``creation" operators \cite{m}
are in fact only factorization operators, since they do not commute to the
identity. Thus, they do not reduce to the common pair of
conjugated oscillator operators ($a, a^{\dagger}$)
in the isospectral infinite limit but merely to
nonlinear product type operators ($a^{\dagger}a^2$, $(a^{\dagger})^2a$).
They connect the discrete part of the spectrum to the continuum part and
vice versa, and
in fact this is the reason why the old isospectral construction is directly
related to inverse scattering methods \cite{n}.
In group theory terminology, the factorization operators mix Fock
representations (bound states) and non-Fock ones (continuous spectrum).
Such a mixing leads, in the isospectral infinite
limit, to the aforementioned (cubic) nonlinear operators.
The true nature of such operators is not yet
very clear, and depend on the particular case under consideration.
The limit, itself, suggests a possible connection by means of the
isospectral parameter
\break{0.7in}
between the linear Schr\"odinger equation and the cubic nonlinear one.
I mention that recently, Andrianov {\em et al} \cite{acdi}
related the bound and continuous
part of a spectrum by means of local q-deformed oscillators.
The relation between the isospectral parameter and
the q-parameter
(q-self-similarity) is an interesting open issue. Solitonic operators,
such as Mielnik ones, may be quite useful in more concrete applications,
as they reflect the connection between the bound and unbound spectra,
which cannot be avoided.
\bigskip
\bigskip

I thank A. Khare for a previous correspondence.

This work was supported in part by CONACyT 4868-E9406.

\end{document}